\documentclass[12pt,dvips]{article}

\usepackage{amsmath,amssymb,exscale}
\usepackage{array,multicol}
\usepackage{afterpage,float,flafter}
\usepackage{epsfig,rotating,pifont}
\usepackage{cite}
\input{epsf}
\@addtoreset{equation}{section} \makeatother

\begin{document}

\vskip 1cm

\begin{center}
{\Large \bf  Black Holes with Flavors of Quantum Hair?}

\vskip 1cm {Gia Dvali\footnote{\it  email:  dvali@physics.nyu.edu}}

\vskip 1cm
{\it Center for Cosmology and Particle Physics, Department of Physics, New York University, New York, NY 10003}\\
\end{center}

\date{}

\begin{abstract}

 We show that  black holes can posses a long-range quantum hair of  super-massive tensor fields,
  which can be detected by  Aharonov-Bohm  tabletop interference experiments,  in which a 
 quantum-hairy black hole, or a remnant  particle,  passes through the loop of a magnetic solenoid.    
 The long distance effect does not decouple  for an arbitrarily high mass  of the hair-providing  field.  
 Because Kaluza-Klein and String theories contain infinite number of massive tensor fields, we study black holes with quantum Kaluza-Klein hair. We show that in five dimensions  such a black hole can be interpreted as a string of  `combed'  generalized magnetic monopoles, with their fluxes confined  along
 it.  For the compactification on a translation-invariant circle, 
this substructure uncovers  hidden flux conservation and quantization of the monopole charges, which constrain the quantum hair of the resulting four-dimensional black hole. For the spin-2 quantum hair  this result is somewhat  unexpected, since the constituent `magnetic' charges   
have no `electric' counterparts.  
  Nevertheless, the information about their quantization is encoded in singularity.

\end{abstract}

\newpage
\renewcommand{\thepage}{\arabic{page}}
\setcounter{page}{1}

\section{Introduction}

 It is well known\cite{nohair,nohair1}  that, in accordance to standard no-hair theorems \cite{nohair3, nohair, nohair1}, 
  black holes cannot maintain any  time-independent classical hair 
 of massive  fields, in particular,  a hair of a massive spin-2 meson. 
  In \cite{quantum}  we have shown that although classical hair is absent, 
 nevertheless black holes can posses a quantum mecanical hair 
 under the massive spin-2 field. 
  This hair can be detected at infinity  by means of the 
 Aharonov-Bohm\cite{AB} effect.  The similar statement is true about the massive antisymmetric Kalb-Ramond
 two-form.  
 Phenomenologically,  the key  fact is the {\it non-decoupling} phenomenon, due to which  the large distance effect 
 survives for an arbitrarily high  mass of  the hair-providing field. 
 
 Interestingly,  the microscopic nature of the  detector string is completely unimportant, provided action contains certain boundary terms.
We shall show, that the detecting string in question can even be an ordinary magnetic solenoid. 
 The black holes (or their hair-carrying remnants) will then experience an 
 Aharonov-Bohm type phase shift,  when encircling such a solenoid.  The effect, that could in principle be  detected in the tabletop setup. 

 It is well known that Kaluza-Klein (KK) and string theories contain infinite number of the massive tensorial gauge fields, which are effectively in the Higgs phase.  Thus, in such theories one may  expect the black holes to carry a variety of quantum-mechanically-detectable conserved charges. 
 It is therefore important to understand the fundamental nature of these charges.

 As a step in this direction, we study black holes with quantum hair  under  Kaluza-Klein excitations. 
 We show that from five dimensions such a black hole can be interpreted 
 as a string of  `combed'  generalized magnetic  monopoles, with their  fluxes confined along it. 
  The monopoles in question,  of course, are not electromagnetic and  belong to the corresponding 5D tensor field  theory. 
  
  The physical picture can be described in analogy with the ordinary Aharonov-Bohm story. 
 If we think of the direction of the magnetic solenoid, as the `extra' dimension, then magnetic flux is 
 confined in the solenoid and directed into the extra dimension, whereas in effective low dimensional theory, on $2+1$-dimensional slices,  the hair is purely quantum.  In the analogous way, 
 in our case the generalized magnetic flux is directed along the solenoid in fifth 
 dimension, whereas in effective $3+1$-dimensional theory the hair is purely quantum.  
  
 This representation  uncovers conditions of underlying flux-conservation and/or quantization of magnetic charges,  which strongly  constrain  properties of quantum hair of the 4D black hole. 
 
  For antisymmetric form theories, the above constraint  is relatively easy to understand. As we shall see, in the presence of the quantum hair, these theories usually involve both magnetic and electric sources.  In particular, for 
  5D massless antisymmetric Kalb-Ramond field,  with the usual minimal `electric'  
 coupling to a string, we find the following.  For the compactification on a circle (assuming 
 unbroken translational invariance along it) the information about any measurable quantum hair is entirely encoded into a zero mode.
  Of course, the detectability of the hair persists even if the zero mode becomes arbitrarily massive,
  due to the  Higgs effects. However, this finding is important in the sense that whole KK 
 tower delivers a single measurable charge in 4D. 
  
   For the three-form theory, the story is different, because the probes of the quantum hair in 5D 
   are 2-branes,  as opposed to  strings. The minimal coupling of such a three-form  does not play the 
   role of the electric  counterpart for the underlying generalized magnetic  monopoles. 
   Derivation of  the constraint is less straightforward, and follows from the requirement of 
   flux conservation of the underlying two-form gauge parameter.

  For the spin-2 case, the situation is even more subtle, because no analog of the antisymmetric form 
  electric coupling exists.  However, as we shall see,  for copmactification on a  translation-invariant  circle,  some hidden quantization condition nevertheless 
  follows, which restricts the observable properties of quantum hair, although in a way different from 
  the Kalb-Ramond case. 
  
  For spin-2 field, existence of such quantization is somewhat unexpected, since the constituent magnetic monopoles have no 
  electric counterparts, and  moreover, they don not admit invariant definition in terms of flux integrals 
  constructed out of the spin-2 field.  
  
   The resolution 
  of this seeming puzzle is in the nature of the black hole singularity, which somehow encodes the information  about the electric sources, although no such sources  are present in the effective theory.  
 
 In certain sense, in spin-2 case the `magnetic' constituents  of the string are never real. 
  So, for spin-2 quantum hair, 
 the monopoles are best to be viewed as an auxiliary construct, convenient
 for constraining the quantum hair.  In this respect they should not be confused with gravitational 
 magnetic charges discussed in\cite{gravimagnetic}. 
 
  Finally,  in case when the background does not respect the translation-invariance in the extra coordinate, 
  the story changes, since the probe strings can be restricted  to the fixed surfaces, and loop of loops cannot be  arbitrarily  deformed.  Also the nature of the possible probes and the boundary terms  may change altogether.  For example, the spin-2 quantum hair can now be probed by 2-branes. 
 
 For phenomenological perspective, the most interesting cases, of course, are when the probe 
 of the quantum hair can be an ordinary magnetic solenoid, and we shall make emphasis on understanding  such possibilities.

  Before proceeding, we should stress some important complementary work.  A different type of the quantum mechanical hair, 
  under the {\it discrete gauge symmetries}, was discovered by Krauss and Wilczek  in 
  \cite{discrete}, and was further studied in\cite{discrete1}.  Such a hair can exist  when a gauge 
  symmetry group is Higgsed down to a discrete subgroup, and  can be detected  by the Aharonov-Bohm effect if the black hole encircles a cosmic string that carries the flux of a massive gauge field\cite{ABmassive}.

  Also, the authors of \cite{pseudo} discovered black holes with the hair under the massless axion field, and showed that it can be detected by the Aharonov-Bohm effect, if  the black hole encircles the 
 axionic  cosmic string.  Below, we shall show that for compactifications on a translation-invariant 
 circle, the measurable quantum charges of the whole Kalb-Ramond KK tower effectively reduce to this one. 
  
   Apart of other fundamental differences, the phenomenological peculiarity  of the quantum-mechanical hair 
 found in \cite{quantum} and in the present work, is that even for an arbitrarily high mass 
 of a hair-providing field, the hair  can in principle be detected in a tabletop setup by an  ordinary magnetic solenoid, without need of any  super-heavy cosmic strings.

 \section{The Essence of the Quantum Hair}
 
  Reduced to its bare essentials, the idea of the quantum hair of \cite{quantum} can be represented in the 
  following way.   Consider a vector (one-form)  $\mathcal{A}_{\mu}$, which has the following boundary 
  interaction with a string
  \begin{equation}
\label{vectorandstring}
q\int dX^{\mu} \wedge dX^{\nu} \, F_{\mu\nu} \, + \, {\rm action~for}~ \mathcal{A}_{\mu},
\end{equation}
 where $F_{\mu\nu} \, = \, \partial_{\mu}\mathcal{A}_{\nu} \, - \, \partial_{\nu}  \mathcal{A}_{\mu}$, and 
 $X^{\mu}$ are the string target space coordinates. $q$ is a constant. Since we are interested in large distance effects, the microscopic nature of the string is unimportant.
 
  The above interaction is a boundary term, and is irrelevant classically. However, quantum- mechanically it leads to the observable effects of topologically-nontrivial configurations.   
For example, consider a situation when $F_{\mu\nu}$ has a form of  Dirac's magnetic monopole

\begin{equation}
\label{dmonopole}
F_{ij} \, = \, \mu \, \epsilon_{ijk} {x^k \over r^3}, ~~~~~F_{0j} \, = \, 0,
\end{equation}  
where $x_j$ are space coordinates and $\mu$ is the magnetic charge. 

Imagine for a moment that  $\mathcal{A}_{\mu}$ is a vector potential of some $U(1)$ gauge group. Then, 
(\ref{dmonopole}) tells us that even an $U(1)$-neutral string could detect such a monopole 
by the Aharonov-Bohm experiment, in which the string loop lassoes the monopole. 
The phase shift resulting from such an experiment would be 
\begin{equation}
\label{shift1}
{\rm phase~shift}\, = \, 4\pi \mu q.
\end{equation}  
This  observation is interesting {\it per se}, since it tells  us that  the boundary effects could detect an otherwise completely decoupled  magnetic type charges, the knowledge that one could attempt to use in 
the laboratory searches for such charges.  However, the situation that we are interested in 
is more dramatic.  We wish to consider the case in which the `magnetic'-type hair is simply
classically unobservable. 
 
  This is the case, if $\mathcal{A}_{\mu}$ is not a massless $U(1)$-field, but a component of 
  a massive tensorial  gauge field (e.g., spin-2).  I such a case, the rest of the action (the first term in (\ref{vectorandstring}) automatically has it)  should have 
  a higher gauge symmetry, under  
\begin{equation}
\label{gaugeA}
\mathcal{A}_{\mu} \, \rightarrow \, \mathcal{A}_{\mu} \,  - \, \xi_{\mu}, 
\end{equation}
where $\xi_{\mu}$ is an arbitrary regular vector. 
Because of this higher gauge symmetry,  from the point of view of the massive tensorial theory
the configuration (\ref{dmonopole}) is {\it locally-pure-gauge}, and thus the hair becomes 
{\it quantum}.  Classically, all the gauge invariant observables, including the energy momentum tensor, 
vanish locally everywhere outside the singularity at $r=0$, and are impossible to detect by any 
local classical measurement.  Due to this fact, the black holes can be endowed with such a hair, without 
any conflict  with the classical no-hair theorems\cite{nohair3,nohair,nohair1}. 

 The gauge symmetry (\ref{gaugeA}) in case of the spin-2 massive field is realized through the 
 Pauli-Fierz action, which in the presence of boundary terms can be written as \cite{quantum}
\begin{equation} 
\label{total1}
\int_{3+1}\hat{h}^{\mu\nu} \mathcal{E}^{\alpha\beta}_{\mu\nu} \hat{h}_{\alpha\beta}\, 
\, - \, m^2\,  \left ((\hat{h}_{\mu\nu} \, +  m^{-1}\partial_{\{\mu}\mathcal{A}_{\nu\}})^2 \, - \, {1\over 2} (\hat{h} \, + \, 2 m^{-1} \partial^{\alpha}\mathcal{A}_{\alpha})^2 \,  \right)\, + \, 
\end{equation}
\begin{equation}
\nonumber
+\, q \,\int_{1+1}  dX^{\mu}\wedge dX^{\nu} \, F_{\mu\nu},
\end{equation}
where the first term is the usual linearized Einstein's action. 
The invariance under (\ref{gaugeA}) is maintained by the corresponding shift 
\begin{equation}
\label{gaugeh}
\hat{h}_{\mu\nu} \, \rightarrow \hat{h}_{\mu\nu}  \, +  \,m^{-1} \, ( \partial_{\mu}\xi_{\nu} \, + \, \partial_{\nu} \xi_{\mu}),
\end{equation}
which remains exact for arbitrarily high $m^2$.
Because of this symmetry, the monopole solution (\ref{dmonopole}) has no classical hair, 
since $\hat{h}_{\mu\nu}$ and $\mathcal{A}_{\mu}$ exactly compensate each other. 

The boundary term (\ref{vectorandstring}), because of its topological nature,  `does not care' whether $\mathcal{A}_{\mu}$
is the part of the massive field, and continues to lead to the same Aharonov-Bohm effect at infinity
even in $m \rightarrow \infty$ limit. 
Thus, the phase shift given by (\ref{shift1}) is insensitive to the mass of the hair-providing field.

\section{Tabletop Detectors?}
  
  As we have seen, the crucial fact for the detectability of the massive quantum hair is  the
  coupling of a two-form $F_{\mu\nu}$, formed out of the compensating  St\"uckelberg field, 
  to a string (\ref{vectorandstring}). This coupling is a boundary term, which is irrelevant for the topologically-trivial 
  configurations. However, for the black holes with the quantum hair it leads to the Aharonov-Bohm 
  type scattering around the string. 
  
  The interesting fact is that the role of the strings in question can be played by the 
  ordinary solenoids, carrying the usual magnetic flux.  The relevant boundary coupling in such a case has the following form 
  \begin{equation}
\label{boundary}
q F_{\mu\nu} \, F_{\alpha\beta}^{(EM)} \epsilon^{\mu\nu\alpha\beta},
\end{equation}
where $F_{\alpha\beta}^{(EM)}$ is the ordinary Maxwellian electromagnetic fields  strength.
For the antisymmetric massive  Kalb-Ramond field the above coupling can also be supplemented by the
coupling $ B_{\mu\nu}  \, F_{\alpha\beta}^{(EM)} \epsilon^{\mu\nu\alpha\beta}$. The latter is not a boundary term, but on the quantum-hair-carrying configuration leads to the similar effect. 
 
In the approximation of an infinitely thin  solenoid, the world-volume current of the solenoid 
becomes a string current, 
\begin{equation}
\label{solenoid}
F_{\alpha\beta}^{EM} \epsilon^{\alpha\beta\mu\nu} ~~~ \rightarrow ~~~ \Phi\, \int dX^{\mu}\wedge dX^{\nu},
\end{equation}
where $X^{\mu}$ are the solenoid coordinates, and $\Phi$ is the total magnetic flux flowing 
in the solenoid.

Thus, if we avaluate the  boundary coupling (\ref{boundary})  on a magnetic-flux-carrying solenoid, it will effectively reduce to
(\ref{vectorandstring}), where the world volume of the string has to be understood as the world volume of 
the solenoid 
\begin{equation}
\label{ABcoupling}
q \Phi\, \int dX^{\mu}\wedge dX^{\nu} \, F_{\mu\nu}.  
\end{equation}
Whenever a black hole or a particle with a quantum charge $\mu$ goes around such a solenoid, 
the wave function of the system acquires the following phase shift
\begin{equation}
\label{pshift}
\Delta S \, = \, 4\pi \mu q \Phi,   
\end{equation}  
just as if the particle had an electric  charge $\mu \, q$, except that in reality there is no such charge, and electric field around the black hole is  zero.  

 The idea of a tabletop experiment that could search for such black holes, 
or their remnant particles, is then simple.  One has to perform the ordinary Aharonov-Bohm experiment 
and look for the deviations from the result predicted by electrodynamics.  For instance, one 
could perform an experiment on the neutral particles, which locally create no electromagnetic field. 
For such particles the ordinary Aharonov-Bohm effect should be absent, unless the 
particles in question carry the quantum hair under some massive fields.  Non-trivial outcome for such an 
experiment would be a signal of a hidden quantum hair of a  particle.

  It is well known, that Kaluza-Klein or String theories (or even QCD) contain infinite tower of massive tensorial  gauge fields. From the 4D effective field theory point of view,  these theories incorporate 
 an infinite number of gauge symmetries. Thus, we may expect  in such theories black holes to posses 
 large number of quantum charges.

   From the point of view of the effective 4D theory, the 
 relevant part of the action coming from  the linearized five dimensional graviton 
 can be represented in the following form
\begin{equation} 
\label{total}
 \sum \, \Big\{
\int_{3+1}\hat{h}^{(m)\mu\nu} \mathcal{E}^{\alpha\beta}_{\mu\nu} \hat{h}^{(m)}_{\alpha\beta}\, 
\, - \, m^2\,  \left ((\hat{h}_{\mu\nu}^{(m)} \, + \, m^{-1} \partial_{\{\mu}\mathcal{A}_{\nu\}}^{(m)})^2 \, - \, {1\over 2} (\hat{h}^{(m)} \, + \, 2 m^{-1} \partial^{\alpha}\mathcal{A}_{\alpha}^{(m)})^2 \,  \right)\, + \, 
\end{equation}
\begin{equation}
\nonumber
+\, q(m)\,\int_{1+1}  dX^{\mu}\wedge dX^{\nu} \, F_{\mu\nu}^{(m)}
    \Big\},
\end{equation}
 which describe an usual KK tower 
of  massive spin-2 particles with five degrees of freedom.   From effective 4D perspective one expects the black holes  in this theory to be labeled by an (infinite) variety of the additional quantum mechanical spin-2 charge $\mu(m)$, which are 
 undetectable classically, but are observable quantum mechanically.  What we wish now to understand is what are the constraints on these charges coming from underlying fundamental theory.

 \section{Quantum  Kaluza-Klein Hair}
 
  We shall now try to construct  black holes with the quantum  Kaluza-Klein  hair.   
   For simplicity of demonstration, we shall consider a single extra dimension on an asymptotically  flat background.    
   We shall try to first understand such {\it quantum-hairy}  black holes directly in five dimensional language, and 
   then go to the compactified version of the theory. 
  Because our arguments are topological, the most important point is to understand the 
  behavior of the solutions at infinity, where linearized gravity is a good approximation.  
  Our starting point, thus, will be an equation for the massless graviton in five-dimensions, far away from the sources. This equation has the following form 
\begin{equation}
\label{einstein}
 \square h_{AB} \, - 
\, \square \eta_{AB} h  \, - \, \partial^{C}\partial_{A} h_{CB} \, -\, 
\partial^{C} \partial_{B} h_{CA} \, + \, 
\eta_{AB} \partial^{C}\partial^{D}h_{CD}\, + \, \partial_{A}\partial_{B} h\, = \, 0,
\end{equation}
where the capital Latin indexes are the 5D ones.  The four-dimensional indexes we shall denote by the
lower case Greek letters.  Because of the five-dimensional gauge invariance, 
\begin{equation}
\label{gaugeh}
h_{AB} \, \rightarrow h_{AB}  \, +  \, \partial_{A}\xi_{B} \, + \, \partial_{B} \xi_{A},
\end{equation}
where $\xi_{A}$ is an arbitrary regular five-dimensional vector (one-form), the equation (\ref{einstein}) 
automatically has a {\it locally-pure-gauge} solution, for which  
\begin{equation}
\label{gaugesolution}
h_{AB} \, =  \, \partial_{A}\xi_{B} \, + \, \partial_{B} \xi_{A}.
\end{equation}
We, however, wish to choose $\xi_A$ to be topologically non-trivial and therefore globally not removable by the pure gauge transformation. For this we first choose, 
\begin{equation}
\label{xi}
\xi_5 \, = \, 0, ~~~{\rm and} ~~~\xi_{\mu}(y,x) \, = \, f(y)\, A_{\mu}(x), 
\end{equation}
where $y$ stands for the $5$-th coordinate. 

 Following \cite{quantum}, we shall now choose $A_{\mu}(x)$ in a form that for an ordinary  
electromagnetic vector potential  would give a Dirac magnetic monopole.  
  For instance, we can choose in the spherical coordinates,  
 \begin{equation}
\label{amonopole}
A_{\phi} \, =  \, {1 \, - \cos \theta \over r \sin \theta},~~~A_{\theta} \, =\, A_r\, = \, 0, 
\end{equation}
corresponding to the spherically symmetric radial magnetic field
\begin{equation}
\label{magnet}
\vec{\mathcal{M}}\, = \,  {\vec{r} \over r^3}.
\end{equation} 
In this gauge, the Dirac string coincides 
with the negative $z$ semi-axis. Just as in the case of a single massive spin-2 field studied in \cite{quantum},   unobservability of the  Dirac string is guaranteed  
by the fact that there are no electrically charged particles under $A_{\mu}$,  since the latter is not 
an $U(1)$ gauge field. 

 As in the case of the Dirac magnetic monopole, we can describe the above configuration without any reference to the string singularity.  We can define the smooth  vector potentials  
on the two hemispheres in the following way  \cite{hemispheres}
\begin{eqnarray}
A_{\phi}^{U} & = &  {1 \, - \, \cos \theta \over r  \sin\theta}~~~~~~~~~0 \, < \, \theta \, < {\pi \over 2}  \\
A_{\phi}^{L} & = & - \,  {1 \, + \, \cos \theta \over r  \sin\theta}~~~~~~ {\pi\over 2}  \, < \, \theta \, < \pi. 
\end{eqnarray}
In the same time we define $h_{AB}^U, h_{AB}^L$ according to (\ref{gaugesolution}). 
Because $A^U_{\mu}$ and $A^L_{\mu}$ at the equator differ  by a single-valued gauge transformation, 
so do $h^U$ and $h^L$
and thus, they  describe the same physics.  In terms of  $A_{\mu}$ the components of the graviton 
have the following form 
\begin{eqnarray}
\label{hform}
 h_{\mu\nu} (x,y) \, &=& f(y) \,  (\, \partial_{\mu}A_{\nu}(x)  \, + \, \partial_{\nu} A_{\mu}(x))
 \\
 \label{h2}
 h_{\mu 5} (x,y) \, &=& \, f'(y) A_{\mu}(x)
 \\
 \label{h3}
 h_{5 5} (x,y) \, &=& 0
\end{eqnarray}
where $f'(y) \equiv \partial_y f(y)$. 
 What is the physical meaning of the above solution? 
 

 We shall show below that it can be interpreted as one dimensional array of `combed' 5D 
 gravitational magnetic  monopoles.  Under `combed' we mean a monopole
 whose entire magnetic hair is confined  into a one dimensional  flux tube.  The field outside such a tube
 is pure gauge.  In ordinary gauge theories, either Dirac or 't Hooft-Polyakov monopole flux gets confined into a string, whenever 
 $U(1)$  gets Higgsed.  The magnetic field of the unconfined monopole is measurable at large distances.  Gravitational monopoles that we are talking about are peculiar in the sense that they are always locally-pure-gauge configurations.   
 
  Because of the same important difference (that $A_{\mu}$ is not an Abelian gauge field), the difference 
  from the Dirac magnetic monopole is that  configuration  (\ref{hform}\, - \, \ref{h3})  has an exactly zero energy everywhere 
  except at one point $r=0$.  In contrast, for the Dirac monopole  the energy diverges  as $1/r$, for 
  $r \rightarrow 0$. 

 Because of the physical string singularity at $r=0$, the solution cannot be pure gauge everywhere, 
 and metric near $r=0$ should also contain a classically-measurable  component.  By its symmetries and the toplogical structure the classical  part of the metric must represent the one of a black string with non-uniform longitudinal density (or traveling longitudinal waves).   Examples of solutions  for non-uniform density  black strings are known  \cite{blackstring}.  Whether the field configuration of our interest 
 matches any of these known results, is unclear.  Nevertheless, we shall sometimes refer to the singularity  at the origin as the `black string', due to its one dimensional nature and the expected topology.  
  
   Close, to the black string the non-linear effects are important, and our locally-pure-gauge form also will be completed by the fully non-linear gauge 
 transformation. We shall not attempt to find the explicit form here, since  all we need is to know that 
 such completion must take place, because of the non-trivial topology at infinity. 
  The non-trivial topology comes from the fact that we are mapping the string of the magnetic monopoles 
 with the topology $S_2 \times  R_2 \times R_1$ on a black string metric of the same topology.   

 We also do not know the stability properties of the singular black string in uncompactified  5D space. 
However, since at the end of the day,  we are interested in compactified  version in which the 
5D black string is entirely `swallowed' by an interior of a 4D black hole, which (by default)  extends 
across the entire 5D bulk,   the 5D stability properties have limited relevance for our analysis.
What is important for us is  to trace the effective 4D properties measurable at 4D infinity. 
 Into whatever states, with measurable quantum numbers, the 5D black string could potentially decay, 
these quantum numbers will continue to belong to the 4D black hole. 

\section{Compactification} 

 Since  we are interested in the 4D black holes, we shall now compactify the $y$-coordinate on a circle  
of radius $R$.  With the black string wrapped on it.   The black string then becomes a black hole. 
 By the classical no-hair theorems\cite{nohair3, nohair, nohair1}, the 4D metric outside the horizon should not contain any non-zero KK
 admixture, observable classically.  
 However,  we are interested in the long range properties 
 of the quantum mechanical hair.  In order to trace them, let us consider a dimensionally reduced version of the equation for a five-dimensional graviton
 (\ref{einstein}).  We first  assume the usual KK anzats  
 \begin{eqnarray}
\label{hmonopole}
 h_{\mu\nu} (x,y) \, &=& \sum \, b^{(m)}(y) \hat{h}_{\mu\nu}^{(m)}
 \\
 h_{\mu 5} (x,y) \, &=& \sum \, b^{(m)'}(y) \, A_{\mu}^{(m)}(x)
 \\
 h_{5 5} (x,y) \, &=& \sum \, b^{(m)''}(y) \, \phi^{(m)}(x),
\end{eqnarray}
where,  $b^{(m)}$ is the complete set of  the usual KK  harmonic functions, satisfying  
 \begin{equation}
\label{bfunctions}
b^{(m)}(y)'' \, = \, - \, m^2 \, b^{(m)}(y). 
\end{equation} 
Then,  going to the new variables, 
\begin{equation}
\label{aphysical}
\mathcal{A}_{\mu}^{(m)} \, \equiv \, A_{\mu}^{(m)}(x) \, + \, {1 \over 2}  \partial_{\mu}\phi^{(m)}(x),
\end{equation} 
the equation (\ref{einstein}) splits in the set of the following equations,
  \begin{equation}
\label{pfs}
 \mathcal{E}^{\alpha\beta}_{\mu\nu} \hat{h}_{\alpha\beta}^{(m)}\, 
\, - \, m^2\,  (\hat{h}_{\mu\nu}^{(m)} \, - \, \eta_{\mu\nu} \hat{h}^{(m)}  \, + \partial_{\mu}\mathcal{A}_{\nu}^{(m)} \, + \, \partial_{\nu} \mathcal{A}_{\mu}^{(m)} \, - \, 2 \eta_{\mu\nu} \partial^{\alpha}\mathcal{A}_{\alpha}^{(m)}) \,  = \, 0
\end{equation}
and
 \begin{equation}
\label{Aequ}
\partial^{\mu} F_{\mu\nu}^{(m)} \, +  \, \partial^{\mu}  (\hat{h}_{\mu\nu}^{(m)}\, - \, \eta_{\mu\nu} \hat{h}^{(m)}) \, = \, 0,
\end{equation}
where $F_{\mu\nu}^{(m)} \, = \, \partial_{\mu}\mathcal{A}_{\nu}^{(m)} \, -\, \partial_{\nu}\mathcal{A}_{\mu}^{(m)}$ and 
\begin{equation}
\label{einstein4}
\mathcal{E}^{\alpha\beta}_{\mu\nu} \hat{h}_{\alpha\beta}^{(m)}\, \equiv \, \square \hat{h}_{\mu\nu}^{(m)} \, - 
\, \square \eta_{\mu\nu} \hat{h}^{(m)}  \, - \, \partial^{\alpha}\partial_{\mu} \hat{h}_{\alpha\nu}^{(m)} \, -\, 
\partial^{\alpha} \partial_{\nu} \hat{h}_{\alpha\mu}^{(m)} \, + \, 
\eta_{\mu\nu} \partial^{\alpha}\partial^{\beta}\hat{h}_{\alpha\beta}^{(m)}\, + \, \partial_{\mu}\partial_{\nu} 
\hat{h}^{(m)}
\end{equation}
is the linearized Einstein's tensor, and as usual $\hat{h}^{(m)} \equiv  \eta^{\alpha\beta}\hat{h}_{\alpha\beta}^{(m)}$. Notice, that $\mathcal{A}_{\mu}^{(m)}$ has a different dimensionality than in (\ref{total}).

  As usual, after compactification  $m$ becomes quantized  
in units of $1/R$. 
The above system has the infinite number of gauge invariances (one per each KK level)  
\begin{equation}
\label{gauge}
\hat{h}_{\mu\nu}^{(m)} \, \rightarrow \hat{h}_{\mu\nu}^{(m)}  \, +  \, \partial_{\mu}\xi_{\nu}^{(m)} \, + \, \partial_{\nu} \xi_{\mu}^{(m)},
~~~~~~~\mathcal{A}_{\mu}^{(m)} \, \rightarrow \, \mathcal{A}_{\mu}^{(m)} \,  - \, \xi_{\mu}^{(m)}.
\end{equation}
Thus, as seen from the four dimensional point of view, $\mathcal{A}_{\mu}^{(m)}$  act as St\"uckelberg  
fields for the massive spin-2 KK states. From this point on, we shall repeat the construction of
\cite{quantum}. 

There is a topologically-nontrivial spherically-symmetric configuration, for
 which $\hat{h}_{\mu\nu}^{(m)}$ has a pure gauge form  locally-everywhere (away from the black hole singularity).  This is the configuration for which  $F_{\mu\nu}^{(m)}$ has a form of the magnetic field of a Dirac monopole
 placed at the center, with
 $\hat{h}_{\mu\nu}^{(m)}$ satisfying 
 \begin{equation}
\label{hmonopole}
 \hat{h}_{\mu\nu}^{(m)} \, = - (\, \partial_{\mu}\mathcal{A}_{\nu}^{(m)} \, + \, \partial_{\nu} \mathcal{A}_{\mu}^{(m)}), 
\end{equation}
with $\mathcal{A}_{\mu}^{(m)} \, = \, \mu(m) A_{\mu}(x)$, and $A_{\mu}(x)$  having the form (\ref{amonopole}).  This configuration can be simply translated in 
(\ref{hform}\, - \, \ref{h3}) in terms of the KK expansion of the function $f(y)$, 
\begin{equation}
\label{fkk}
f(y) \, = \, \sum  b^{(m)}(y) \mu(m). 
\end{equation}
Thus, the monopole charges $\mu(m)$ of individual KK-s are simply KK-harmonics of  the function $f(y)$.  

Note, that all the  gauge invariant  
observable massive spin-2 fields, 
 \begin{equation}
\label{hKK}
h_{\mu\nu}^{(m)} \, \equiv  \,  \hat{h}_{\mu\nu}^{(m)} \, +  (\, \partial_{\mu}\mathcal{A}_{\nu}^{(m)} \, + \, \partial_{\nu} \mathcal{A}_{\mu}^{(m)}), 
\end{equation}
are identically zero on this configuration, in full accordance  with the Bekenstein's classical result 
\cite{nohair}. 

\section{Interpretation in Terms of 5D Gravitational `Monopoles'} 

 We shall now try to interpret our `black string' solution as an array of the point-like 5D objects, 
with their magnetic-type fluxes confined in tubes oriented along $y$-axis.  We shall refer to these
`constituents'
as gravitational magnetic monopoles,  although as we shall see they do not posses any detectable 
magnetic field. 

 Possibility of such interpretation is important because of the following reason.  
As discussed above,  in effective 4D theory the detection of the quantum hair requires a boundary coupling (\ref{vectorandstring}). In 5D theory, this coupling is embedded into the boundary coupling of the following form 
 \begin{equation}
\label{5dstring}
q_5 \, \int  dX^{A}\wedge dX^{B} \, \partial_{[A} \partial^C h_{CB]}, 
\end{equation}
 where $q_5$ is the coupling constant and  $X^{A}$ are the string coordinates in  5D.
This boundary term represents an `electric'-type coupling between the conserved string current 
and a two-form $X_{AB} \equiv   \partial_{[A} \partial^C h_{CB]}$, which is the lowest possible antisymmetric two-form constructed out of $h_{AB}$.  We therefore wish to understand whether 
the 5D coupling (\ref{5dstring})  satisfies any consistency quantization condition.   

 Usually, the quantization of the electric charges follows from the existence of the magnetic ones. 
We thus wish to understand if there are charges that can be magnetic with respect to (\ref{5dstring}). 
  
   In 5D in order to define a magnetic monopole configuration, one needs an existence of an
antisymmetric  two-form  (call it  $B_{AB}$) for which we can define a three-form field strength \begin{equation}
\label{FB}
H_{ABC}\, \equiv\, \partial_{[A} B_{BC]} \, \equiv \, dB,
\end{equation}
where $d$ is the exterior derivative. 
The magnetic monopole then represents a configuration for which there is a non-zero flux through 
a three-sphere $S_3$   
 \begin{equation}
\label{5dflux}
 \int_{S_3} dX^{A}\wedge dX^{B}\wedge dX^{C} \, H_{ACB}\, = \, 2\pi^2\, \bar{\mu}. 
\end{equation}
In particular, for a static spherically symmetric case $H_{ABC}$ has a Freund-Rubin\cite{fr} type form 
\begin{equation}
\label{smagnet} 
H_{ABC} \, = \, \bar{\mu} \, \epsilon_{ABCD0} \, {x^D \over r^4}.
\end{equation}
For an elementary two-form $B_{AB}$ field, such a  monopole is a solutions of 
equations of  motion\cite{5dmonopoles}. This will play an important role when we shall discuss 
a quantum hair of such an elementary field. 

 However, irrespective of whether $B_{AB}$ is an elementary field or a composite operator, the 
 existence of magnetic monopoles leads to quantization of any electric type coupling of
 $B_{AB}$  
 to any conserved source $J^{AB}$,  
 \begin{equation}
\label{qcoupling}
 e B_{AB}J^{AB}. 
\end{equation} 
Quantization follows from the requirement of non-observability of the Dirac string, and is a generalization 
of the usual Dirac's quantization condition, which (in proper normalization) demands \cite{5dmonopoles}
\begin{equation}
\label{quant}
\bar{\mu}\, e\,  = \, {n \over \pi}.
\end{equation} 

 In trying to apply similar quantization arguments to the coupling (\ref{5dstring}), we immediately run  
 into the following  problem. Because $X_{AB}$ is itself an exterior derivative, any field strength evaluated out 
 of it is automatically zero.
 
  Due to this property, the coupling (\ref{5dstring}) seems unconstrained, since the system 
 admits no magnetic monopoles with invariant definition of the charge, in terms of a surface integral
 of a three-form field strength. 
 
  Interestingly  this statement is only half-true. Although, it is true that no 5D-invariant definition (in terms of a local  antisymmetric form constructed out of $X_{AB}$) of the monopole charge is possible, 
  nevertheless some 5D monopole type objects
  can be identified, which can constrain the possible form of  quantum charges $\mu(m)$. 
  
In order to find such monopoles, we introduce an {\it auxiliary} antisymmetric form $B_{AB}$, and 
require $h_{AB}$ to have the form defined from the following equation 
\begin{equation}
\label{hB}
\square h_{AB} \, = \, \partial_A\partial^CB_{CB} \, + \, \partial_B\partial^CB_{CA}.
\end{equation}     
For an arbitrary $B_{AB}$ such an anzats is pure-gauge, and outomatically solves the equation 
for $h_{AB}$.  Now, we are free to choose $B_{AB}$ in the form of a magnetic monopole with a non-zero flux (\ref{5dflux}).  Notice,  that for the spherically symmetric configuration (\ref{smagnet}), the monopole flux is unobservable neither classically nor quantum mechanically, because $B_{AB}$ is divergence-free and $h_{AB}$  vanishes.  Thus, such a monopole is simply a singular pure gauge configuration. 
However, the flux of a combed monopole becomes observable.  Although, combing of the monopole 
does not change its charge measured by (\ref{5dflux}),  it does nevertheless change the effect measured by (\ref{5dstring}), since the latter coupling admitted no invariant charge definition. 
Putting it crudely, combing of a monopole is not a gauge transformation. 

 Consider now a combed monopole placed at $y=0$ and with all its magnetic flux confined in an 
 infinitely thin string directed along the $y>0$ semi-axis.  For such a monopole, $B_{AB}$ has a pure 
 gauge form locally everywhere except at a singular flux line ($r=0, y>0$). So we have
 \begin{equation}
\label{bflux}
B_{AB} \, = \, \partial_A \xi_{B} \, - \partial_B \xi_A
\end{equation} 
 where, $\xi_A$ has a form given by (\ref{xi}), with 
 \begin{equation}
\label{fforb} 
f(y) \, = \, \bar{\mu} \,\theta(y).
\end{equation}
 The corresponding magnetic field, has the form 
  \begin{equation}
\label{ymagnet} 
H_{ABC} \, = \, \bar{\mu} \, \epsilon_{ABC50} \, \delta^3(x) \theta(y).
\end{equation}
If $B_{AB}$ had an electric coupling of the form (\ref{qcoupling}),  the Dirac quantization condition
(\ref{quant}) would automatically follow.  But since we do not have such couplings, seemingly 
$\mu$ is not quantized. Thus, {a priory} there is no obstruction in creating an arbitrary sequence 
of the combed monopoles along the $y$-axis, with an arbitrary continuous distribution of charges.
Then,   $f(y)$ in (\ref{fforb}) can be chosen to be an arbitrary function\footnote{Except that the line cannot 
terminate.}.  The corresponding form of 
$h_{AB}$ defined from (\ref{hB}) will be the one given by (\ref{gaugesolution}) with 
$\xi_A$ given by (\ref{xi}). 

 We have thus reinterpreted  our initial `black string' configuration as the string of the 
 combed 5D magnetic monopoles.   

 We shall now see, that although there are no electric `duals' for monopoles, and no magnetic monopoles for $X_{AB}$, nevertheless the singularity somehow knows about both, and the 
 function $f(y)$ cannot be arbitrary.  In the other words, there is a hidden quantization condition 
 for the monopole charges, although these monopoles  are pure gauge configurations, 
 with no corresponding electric counter parts.

 \section{Detecting the Quantum Hair and Hidden Quantization Condition}
 
  In order to derive the constraints on the quantum hair of the black string, 
it is instructive to perform the Aharovon-Bohm detection experiment directly in uncompactified 5D version of the theory.  We can do this by lassoing the black string by a probe  string loop coupled 
 to the quantum hair through the boundary term (\ref{5dstring}), which for the given configuration 
 takes the following form
  \begin{equation}
\label{5dstring1}
q_5 \, \int  dX^{A}\wedge dX^{B} \, \square \partial_{[A} \xi_{B]}.
\end{equation}
Assume first  that the probe string stays  in  some $x^5 = y= $ constant  plane.  For such a string $X^5 \, = \, y$, and (\ref{5dstring}) becomes
  \begin{equation}
\label{5dstring1}
q_5 \, f''(y) \int  dX^{\mu}\wedge dX^{\nu} \,  F_{\mu\nu}.
\end{equation}
Evaluating this integral on any closed two-dimensional surface (loop of loops)  surrounding the origin, we get the 
following  phase shift 
   \begin{equation}
\label{shift5}
\Delta S \, = \, 4\pi q_5  f''(y).
\end{equation}
The $y$-dependence of this result makes a little sense, since we could arbitrarily deform the 
closed surface, in order not to lie entirely in a $y=constant$ plane. 
Demanding that the phase shift should be independent under any continuous deformation
of the closed surface that avoids singularity, we are lead to the requirement 
that  $f''(y)\, = \, \mu = \, constant$ everywhere, except possibly at some singular points, where the loop of loops is 
ill defined.   If $\mu q_5 \neq {n \over 2}$, the quantum hair is detectable by the Aharonov-Bohm experiment, but the quantum charges of the whole KK tower effectively sum up  in a single detectable charge $\mu$.  

  For the backgrounds that are translation-invariant  in $y$,
 the only alternative, for making the (\ref{shift5}) consistent is to assume that the  monopole charges are quantized, according to (\ref{quant}),  where $e$ has to be understood as some fundamental unit.  
 In such a case, $f''(y)$ is identically zero, except at the singular points, corresponding to the locations of the (invisible) Dirac strings, that inject magnetic fluxes into the string of monopoles (of course, the 
 junction point coincides with the location of a corresponding monopole on the string). 
An arbitrary loop of loops, that  does not cross the Dirac string, cannot detect any phase shift, and 
in such a case the quantum hair becomes unobservable, even by the Aharonov-Bohm scattering. 

  The fact that one may be lead to the quantization of the magnetic-type  charges in theory without the free electric ones  is {\it per se} rather unexpected.  This feeling  is further strengthened by the fact that, unlike 
 the magnetic charges in $U(1)$ or in high-rank  antisymmetric-form gauge theories,  
 the magnetic charges in question admitted  no invariant definition in terms of any Gaussian  surface integral of  local  three-forms 
 constructed out of  $X_{AB}$, since any such three-form is identically zero.   So in a free state  these
 charges were simply impossible to probe, either classically, or quantum-mechanically.   Yet, existence of the quantum-hair-carrying monople-string solution, can impose a non-trivial quantization condition.
 
  The resolution of this seeming puzzle lies in the singular nature of our combed monopole string solution (\ref{gaugesolution}).  In order to comb the monopole fluxes into an infinitely thin string, the presence  
  of some current sources is necessary. This is analogous  to the situation that, in order to trap an ordinary 
  Maxwellian magnetic flux into a thin tube, one needs an electric-current-carrying solenoid. But, existence of such
  currents implies existence of some electric type charges, that create it. 
   
   We have avoided the explicit introduction of such charges, by declaring the `solenoid' to be 
 infinitely-thin. As a result charges got hidden inside the singularity, and this created a false impression 
 of their absence in the effective  theory. Hence the illusion of unconstrained magnetic charges.
  The singularity, however,  keeps the `memory' about these charges, and responds by the quantization condition (\ref{quant}), which tells us that any short distance physics that would resolve 
 this singularity, inevitably has to employ some electric charges.  The unit $e$ in (\ref{quant}) then, 
 has to be understood as the fundamental unit of latter  electric charges.   
 However, since these  underlying singularity-resolving 
 charges  are not uniquely fixed from effective field theory  point of view, 
 so is the unit of their quatization. 
 
  In general, the effective action does not have to respect the translational invariance in $y$-direction. 
 In such a case, the nature of the constraint may be altered.  For instance, the probe string 
 may be restricted to lie in a fixed $y=constant$ plane. Then, the loop of loops 
 cannot be arbitrarily deformed, and the Aharonov-Bohm effect can only probe a fixed 
 $y$-plane.   In such a case, the restrictions of the function $f(y)$ may be more subtle, since 
 one cannot create inconsistency by simply deforming the string trajectories in
 Aharonov-Bohm  interference experiment.  
 
 In addition, the nature of the probes can change also.  For example,  let $\Sigma$ be some background scalar field, with $y$-dependent expectation value. 
Then, instead of using the boundary coupling to a string  (\ref{5dstring}), we can use the following  boundary coupling 
to a 2-brane, 
 \begin{equation}
\label{2brane}
q_5 \, \int  dX^{A}\wedge dX^{B}\wedge dX^C \, \partial_{[A} (\partial_B\Sigma) \partial^E h_{EC]}, 
\end{equation}
where $X^A$  are the 2-brane coordinates. In the other words, we have created a 
composite two-form $R_{AB} \, \equiv \,   (\partial_{[A}\Sigma) \partial^E h_{EB]}$, and expression 
(\ref{2brane}) is nothing,  but an invariant magnetic  flux of  its field strength, 
\begin{equation}
\label{rform}
\mathcal{R}_{ABC} \, \equiv  \, \partial_{[A} R_{AB]},
\end{equation}
through the world volume  3-surface of the 2-brane.  
 
  We can now perform an Aharonov-Bohm experiment by lassoing  the monopole string by the above 2-brane.  The resulting phase shift is,
 \begin{equation}
\label{shift2brane}
\Delta S \, = \, 4\pi q_5 \int dy  \, \Sigma'(y) \, f''(y),
\end{equation}
and is invariant under a continuous deformation of the 2-brane loop, as long as it does not cross 
the singularity.  The quantum hair is then detectable provided the above expression is not an 
integer times $2\pi$.  The interpretation in terms of the magnetic flux, however, places a 
non-trivial constraint.   The in-flowing `magnetic' flux should either be absorbed by some magnetic sources, 
or continue  to spread  along the black string in $y$-direction.  The net influx to the string can be 
non-zero if $y$-coordinate is non-compact. In such a case there is no immediate restriction on (\ref{shift2brane}), and the phase shift is observable.  For compact $y$, in the absence of sources, 
the total inflow of the flux must vanish. So the phase shift cannot be experienced by  probes that are 
sensitive  to the total influx, but only the ones that may be sensitive to the inflow locally in $y$.  

 We shall now discuss a possible boundary coupling that would enable one to probe
spin-2 KK  hair with an ordinary solenoid, modulo the above restrictions.  
This coupling  is a variant of (\ref{2brane}), and can be written in the following form 
  \begin{equation}
\label{randf}
 \mathcal{R}_{ABC} \, {\bf F}_{DE} \,  \epsilon^{ABCDE}, 
\end{equation}
where ${\bf F}_{DE}(x,y)$ is an Abelian field strength of the five dimensional `pro-genitor' of the usual 
electromagnetic field.  For example, if photon is located at, say,  $y=0$, 
then,  ${\bf F}_{\mu\nu}(x,y) \, = \,\delta(y) F_{\mu\nu}^{EM}(x)$.  Lassoing the monopole string 
by the loop of a magnetic solenoid, with the flux $\Phi$, creats  the phase shift 
 \begin{equation}
\label{shift2brane3}
\Delta S \, = \, 4\pi q_5 \,\Phi \Sigma'(0) \, f''(0).
\end{equation}

  \section{Quantum Hair Under Massive Antisymmetric Fields} 
 
 We shall repeat the similar analysis for the antisymmetric two- and three-form gauge fields, 
 and lift the quantum hair carrying black holes into the string of combed monopole fluxes.

 \subsection{Two-Form Case} 
 
 The Lagrangian of a massless  antisymmetric Kalb-Ramond two-form 
 field $B_{AB}$ in 5D is 
  \begin{equation}
\label{RB}
-\, H_{ABC} \,H^{ABC},
\end{equation}
where $H_{ABC}$ is a three-form field strength defined through (\ref{FB}).
This action is  invariant under  the following gauge symmetry  
\begin{equation}
\label{shiftB}
B_{AB} \, \rightarrow  \, B_{AB}  \, + 
\partial_{[A}\xi_{B]},
\end{equation}
where, as before,  $\xi_{B}$ is an arbitrary regular one-form.  Because of this gage symmetry, the equation 
of motion, 
 \begin{equation}
\label{FBeq}
\partial^AH_{ABC}\, = \, 0,
\end{equation}
 automatically 
admits  a locally-pure-gauge  solution
\begin{equation}
\label{soluB}
B_{AB} \, = \,  
\partial_{[A}\xi_{B]}.
\end{equation}
As in the spin-2 case we wish to make this solution toplogically non-trivial.  We therefore choose $\xi_A$ in 
the form (\ref{xi}), with $A_{\mu}$ having a form of a vector potential of a Dirac magnetic monopole 
given by (\ref{amonopole}).  Again, as in the spin-2 case this topologically non-trivial, but locally-pure-gauge solution amounts to a string-like singularity placed at $r=0$. 
Interpretation of this source in terms of combed $B_{AB}$-magnetic monopoles, is much more 
straightforward than in spin-2 case.   It is known\cite{5dmonopoles} that monopoles are point-like  
solutions of (\ref{FBeq}), with the spherically symmetric magnetic field given by  (\ref{smagnet}). 
Their magnetic charge $\bar{\mu}$ can be  defined as an invariant surface integral (\ref{5dflux}).  
Thus, unlike in spin-2 case the  monopoles in $B_{AB}$ theory carry a `real',  classically-detectable 
magnetic flux.   

 The solution (\ref{soluB}), as in the spin-2 case,  represents an array of combed monopoles with their 
 fluxes confined into strings directed in $y$ direction.  Outside the string the field is pure gauge, 
 and we may attempt to detect   its quantum hair 
 via the following stringy coupling
 \begin{equation}
\label{bstring}
e \int  dX^{A}\wedge dX^{B} \, B_{AB}. 
\end{equation}
This coupling is an electric coupling, and its presence  automatically quantizes
monopole charges  in units of $1/\pi e$. 
 As a result, due to reasons explained in spin-2 discussion, 
the function $f(y) = \bar{\mu}_0 \, + \, n/\pi e$, where $\bar{\mu}_0$ is a constant, and $n$ is an integer 
that can only experience jumps at the singular points corresponding to the junctions of the 
Dirac strings with the string of combed monopoles. 

 Due to this, lassoing the black string
by the probe one,  generates an universal Aharonov-Bohm phase shift
\begin{equation}
\label{ }
\Delta S \, = \, 4\pi \,\bar{\mu}_0 e.
\end{equation}  
The fact that the phase shift is constant means that effectively only the quantum hair of the 
KK zero mode  is probed.  This can be directly seen from a straightforward KK decomposition, which 
shall perform here for completeness. 

 We shall now compactify the $y$-direction on a circle of radius $R$, wrapping  the black string 
on it.  The source now becomes an interior of  a four-dimensional black hole.  As in the graviton case, by classical no-hair theorem all the massive KK fields  must vanish outside the horizon. 
We shall try to trace the quantum hair. 

Assuming the usual KK anzats  
 \begin{eqnarray}
\label{bkk}
 B_{\mu\nu} (x,y) \, &=& \sum \, b^{(m)}(y) \hat{B}_{\mu\nu}^{(m)}(x)
 \\
 B_{\mu 5} (x,y) \, &=& \sum \, b^{(m)'}(y) \, \mathcal{A}_{\mu}^{(m)}(x),
\end{eqnarray}
and rescaling the irrelevant numerical factors, 
the equation (\ref{FBeq}) splits in the set of  equations
  \begin{equation}
\label{eqkkH}
 \partial^{\mu} H_{\mu\nu\alpha}^{(m)}\, 
\, + \, m^2\,  (\hat{B}_{\mu\nu}^{(m)} \, + \, F_{\mu\nu}^{(m)})\,  = \, 0
\end{equation}
and
  \begin{equation}
\label{eqkkH}
  \partial^{\mu}\left ( \hat{B}_{\mu\nu}^{(m)} \, + \, F_{\mu\nu}^{(m)} \right )  = \, 0
\end{equation}
where, $H^{(m)}_{\mu\nu\alpha} \equiv \partial_{[\mu}\hat{B}_{\beta\alpha]}^{(m)}$ and
 as before  $F_{\mu\nu}^{(m)} \, \equiv \, \partial_{\mu}\mathcal{A}_{\nu}^{(m)} \, -\, \partial_{\nu}\mathcal{A}_{\mu}^{(m)}$.

 As in the graviton case, the above system has infinite number of gauge invariances (one per each KK level)  
\begin{equation}
\label{gauge}
\hat{B}_{\mu\nu}^{(m)} \, \rightarrow \hat{B}_{\mu\nu}^{(m)}  \, +  \, \partial_{[\mu}\xi_{\nu]}^{(m)},
~~~~~~~\mathcal{A}_{\mu}^{(m)} \, \rightarrow \, \mathcal{A}_{\mu}^{(m)} \,  - \, \xi_{\mu}^{(m)}.
\end{equation}
Thus, again as seen from the four dimensional point of view, $\mathcal{A}_{\mu}^{(m)}$  act as St\"uckelberg  
fields for massive Kalb-Ramond KK states. 

 Due to this symmetry, there is a topologically-nontrivial spherically-symmetric configuration, for
 which $\hat{B}_{\mu\nu}^{(m)}$ have a pure gauge form  locally-everywhere outside the black hole singularity.  This is the configuration for which  both $F_{\mu\nu}^{(m)}$ and $\hat{B}_{\mu\nu}^{(m)}$ have 
 the form of the magnetic field of a Dirac monopole
 placed at the center, and exactly compensate each other 
\begin{equation}
\label{bmonopole}
\hat{B}_{\mu\nu}^{(m)}  \, = \, - \, F_{\mu\nu}^{(m)} \, =\,  \mu(m) {\epsilon_{\mu\nu} \over r^2},  
\end{equation}
where $\epsilon_{\mu\nu}$ is an induced volume element of a two-sphere enclosing the
Schwarszchild solution.  
That is, we  can choose
the St\"ckelberg fields $\mathcal{A}_{\mu}^{(m)}$ to have the form (\ref{amonopole}).
Notice that the form of each $\hat{B}_{\mu\nu}$ coincides with the massless axion form discovered
in \cite{pseudo}, and generalized to the massive case in\cite{massivepseudo}. 
 
Because, $\hat{B}_{\mu\nu}^{(m)}$ have  a locally-pure-gauge form, 
both the field strength $H$ as well as the gauge-invariant massive KK fields, 
\begin{equation}
\label{bkk}
B_{\mu\nu}^{(m)}  \, \equiv \,  \hat{B}_{\mu\nu}^{(m)}  \, + \,  F_{\mu\nu}^{(m)}, 
\end{equation}
are identically zero everywhere, 
except  the black hole singularity. 

In order to detect the field by the Aharonov-Bohm  effect, one could use the 
coupling (\ref{bstring}). Lassoing the black hole by a string that lies in $y=const$ plane we get for the phase shift  
   \begin{equation}
\label{shift5}
\Delta S \, = \, 4\pi e \int dm \, b^{(m)}(y) \mu(m)\, = \, 4\pi \, e \, f(y). 
\end{equation}
But because, the function $f(y)$ is constant, we have $\mu(m) \, = \, \bar{\mu}_0 \, \delta(m)$, which means that entire contribution comes from the zero mode. 

\subsection{Three-Form Case}

 The Lagrangian  for a massless three-form $C_{ABC}$ in 5D is 
 \begin{equation}
\label{threeaction}
-\,\mathcal{F}_{ABCD}\mathcal{F}^{ABCD}, 
\end{equation}
 where,  $\mathcal{F} \, =  \partial_{[A}C_{BCD]}$ is the invariant field strength.
 Under the gauge transformation, $C$ shifts as 
 \begin{equation}
\label{gaugec}
C_{ABC}  \rightarrow  C_{ABC} \, + \, \partial_{[A}\Omega_{BC]},
\end{equation}
where $\Omega_{AB}$ is a two-form.
Performing a standard KK reduction,  
 \begin{eqnarray}
\label{bkk}
 C_{\mu\nu\alpha} (x,y) \, &=& \sum \, b^{(m)}(y) \hat{C}_{\mu\nu\alpha}^{(m)}
 \\
 C_{\mu\nu 5} (x,y) \, &=& \sum \, b^{(m)'}(y) \, B_{\mu\nu}^{(m)}(x)
\end{eqnarray}
and integrating over the $y$-coordinate, the five dimensional action (\ref{threeaction}) splits 
in the sum of the infinite number of massive three-form actions
\begin{equation}
\label{bcaction}
 L_{C}\, = \,  \sum \, \left ( \, - \, \mathcal{F}_{\mu\nu\alpha\beta}^{(m)}\mathcal{F}^{(m)\mu\nu\alpha\beta}
  \, +  \,   m^2 \, (\partial_{[\alpha}B_{\beta\gamma]}^{(m)}  \, + \, \hat{C}_{\alpha\beta\gamma}^{(m)})^2 \right ),
 \end{equation}
 where 
 $\mathcal{F}^{(m)}_{\mu\alpha\beta\gamma} \, \equiv \partial_{[\mu}\hat{C}^{(m)}_{\alpha\beta\gamma]}$ 
is the invariant field strengths for $m$-th  KK state.  As it is obvious from (\ref{bcaction}), from the point of view of the 4D description each $m\neq 0$ KK represents a three-form that is in the Higgs phase. In this phase the corresponding gauge symmetry is realized through the St\"uckelberg two-forms $B_{\mu\nu}^{(m)}$,
in the following way

\begin{equation}
\label{gaugec}
\hat{C}^{(m)}_{\mu\nu\alpha}  \rightarrow  C_{\mu\nu\alpha}^{(m)} \, + \, \partial_{[\mu}\Omega^{(m)}_{\nu\alpha]},~~~~~~
B_{\mu\nu}^{(m)} \, \rightarrow  \, B_{\mu\nu}^{(m)}  \, - \, \Omega_{\mu\nu}^{(m)}.
\end{equation}
 Notice that also for the action (\ref{bcaction})  the solution (\ref{bmonopole}) goes through undisturbed, since $dB$ vanishes and we can put $C^{(m)}_{\mu\nu\beta} =0$.  So in this case $C^{(m)}_{\mu\nu\beta}$ is not even a pure gauge, but simply zero. 
 
  The 5D interpretation of the above solution is to take 
 \begin{equation}
\label{cpuregauge}
C_{ABC} \, = \, \partial_{[A}\Omega_{BC]},
\end{equation}
where, 
 $\Omega_{AB}$ is the following  two-form,
 \begin{equation}
\label{omega}
\Omega_{\mu\nu}\, = \, 0, ~~~{\rm and}~~~\Omega_{\mu5} \, = \, f(y) \, A_{\mu}(x), 
\end{equation}
where $A_{\mu}(x)$ is taken in the monopole form (\ref{amonopole}). 
Then, $C_{\mu\nu\gamma} \, = \, 0$,  and $C_{\mu\nu 5} \, = \, f(y) F_{\mu\nu}(x)$. 
Since, we are dealing with a three-form, the quantum hair 
can be probed by a 2-brane that sources it,
 \begin{equation}
\label{23brane}
q_5 \, \int  dX^{A}\wedge dX^{B}\wedge dX^C \, C_{ABC}. 
\end{equation}
This is similar to (\ref{2brane}), expect that (\ref{23brane})  has a pure-boundary form only on a given configuration. 
The coupling (\ref{23brane}) gives an invariant magnetic  flux of the three-form, 
\begin{equation}
\label{magomega}
 \partial_{[A} \Omega_{AB]},
\end{equation}
through the world-volume  3-surface of the 2-brane.  The associated  magnetic field vector is radial 
\begin{equation}
\label{bradial}
\mathcal{B}_{i} \, \equiv \, \epsilon_{ijk50}  \partial^{[j} \Omega^{k5]}\, =\, f(y) \, {x_i \over r^3}.
\end{equation}
Thus, there is a radial inflow of the magnetic field to singularity.  The requirement of  flux conservation then demands that, either singularity must incorporate magnetic sources, or the magnetic flux should flow through the singular string in $y$-direction. 

 The latter situation places the following constraint on the 
detectable hair.  First of all,  such a situation cannot be realized  for  the form of $\Omega_{AB}$ 
given by (\ref{omega}), since the $y$-component of the magnetic vector  vanishes. 
$\mathcal{B}_y \, = \, 0$,  because $\Omega_{\mu\nu} \, = \, 0$.   To restore the consistency, we can add
the following locally-pure-gauge contribution to $\Omega$,
 \begin{equation}
\label{omegaextra}
\Omega_{AB} \, = \, \partial_{[A}\xi_{B]}, 
\end{equation}
where, $\xi_A\, = \, \delta_A^{\mu}\, A_{\mu}(x) \, \int_{-\infty}^y \, f(\tau) d\tau$.  The extra contribution is not changing the value of the radial component of the magnetic field (\ref{bradial}), but is producing 
a non-zero $y$-component along the singular  string, 
\begin{equation}
\label{by}
\mathcal{B}_{y} \, = \, \delta^3(r) \, \int_{-\infty}^y \, f(\tau) d\tau, 
\end{equation}
which compensates the radial inflow.

 As in the spin-2 case,   we can now perform the Aharonov-Bohm experiment by lassoing  the singular  string by the above 2-brane.  The resulting phase shift is,
 \begin{equation}
\label{shift2brane2}
\Delta S \, = \, 4\pi q_5 \int dy  \, f(y).
\end{equation}
Since it has a meaning of the integrated magnetic flux,  
it is invariant under a continuous deformation of the 2-brane loop, as long as the latter does not cross 
the singularity.  As said above, demanding its conservation,  the radial magnetic flux has to either  
be absorbed by some sources, or flow along the singular string.  The latter situation 
requires a non-vanishing $y$-component of the magnetic field, given by (\ref{by}). 
The interpretation in terms of the conserved flux, places a non-trivial constraint  on the measured
phase shift (\ref{shift2brane}),  because of the  following reason. 

 On a non-compact space, the net radial inflow of the flux to the string can be non-zero, since the flux can flow along the string to infinity.  If $y$ is compact, however, 
the net influx must vanish, and so will the phase shift.  Thus, in such a situation the hair cannot be probed by the sources that are sensitive to the total flux.  One somehow needs sources that, on one hand, are 
fundamentally restricted to the fixed $y$-planes and, on the other hand, are sensitive to the influx. 
Such a situation,  as  a minimum, requires breaking of the translation-invariance in $y$.

 Finally,  the 5D gauge-invariant coupling, that would enable one to probe a three-form  quantum  hair
 (subject to above-discussed restrictions) with an ordinary magnetic solenoid, can be written in the following form 
  \begin{equation}
\label{candf}
 q_5\, C_{ABC} \, {\bf F}_{DE} \,  \epsilon^{ABCDE}, 
\end{equation}
where, as in the gravity case,  ${\bf F}_{DE}(x,y)$ is an Abelian field strength that includes the usual 
electromagnetic field. 
 For a  magnetic solenoid that is localized at $y=0$, the Aharonov-Bohm phase shift is 
 \begin{equation}
\label{shift2brane1}
\Delta S \, = \, 4\pi q_5 \,\Phi \, f(0),
\end{equation}
where, $\Phi$ is the flux.

 

\section{Conclusions}

  In conclusion, black holes can posses  a quantum  hair 
  under the tensor gauge fields.  Although  in the present work we have limited our analysis by the maximum spin-2 case,  our arguments can be generalized 
 to the higher spin states, that admit St\"uckelberg  description.   

   The quantum  hair can be detected at infinity by the  Aharovon-Bohm  effect, in which a black hole
  or its remnant  passes through a string loop.  The crucial fact is  the {\it non-decoupling} of the
  effect  in the limit of an arbitrarily high mass of the corresponding hair-providing  gauge field.  This gives a prospect of probing a very short distance   physics by large distance observations, possibly even in a
  tabletop  laboratory setup, since the role of the string loop could be played by the usual magnetic solenoid.  
 A black hole or its remnant with the hair, in the presense of such a solenoid, would experience the 
 usual Aharonov-Bohm type interference effect, although they carry no ordinary electric charge.  

   In Kaluza-Klein and String theories,
  where there are infinite number of massive excitations, the black hole quantum hair is also expected to come in different  variety.  We therefore tried to construct  black holes with quantum  KK hair. 
  We found that being uplifted to extra dimensions, such hairy black holes become hairy strings, which can be 
  interpreted as the array of generalized magnetic monopoles with their fluxes confined along the 
  string. The monopoles in question are the monopoles of underlying tensor gauge theory.

 We discovered that, in a simplest  compactification on a circle with exact translation-invariance, 
 this picture  leads  to the quantization of monopole charges,  both for Kalb-Ramond and also for 
 spin-2 massless 5D fields,  which 
 severely constraints the detectable quantum hair of the effective 4D black hole.

  This fact is most unexpected for the spin-2 theory, due to the reason, that constituent magnetic 
  monopoles  in this theory  have no visible electric counterparts.  They also do not admit any invariant 
  definition  of the magnetic charge in terms of the surface integral of an antisymmetric 
  form constructed out of the spin-2 field.  Nevertheless, the quantization follows.  The reason behind this quantization is the string (or the black hole) singularity, which encodes information about the hidden 
  electric type couplings, even though such charges  are absent  in the effective theory.
 At the end, by imposing quatization the magnetic charges make themselves unobservable, and thus 
 only play the role of an auxiliary construction useful for constraining the quantum black hole hair.  
   
 For the backgrounds with broken translation-invariance in the extra coordinate,  the story is different. 
 Both, the nature of the probes of the quantum hair as well as the possible boundary couplings are altered,  and we have investigated  some of these possibilities.   
   
  The fact that the detectable  black hole charges are not necessarily  limited by the ones determined 
 exclusively by the massless gauge field spectrum, should have a direct relevance 
 for the black hole information loss question.  Especially, if the possible new charges can come in infinite number of `flavors'.  The information associated with such charges cannot be lost and must be recovered  after the
 black hole  evaporation.  It is therefore important to understand the extend of possible varieties  of  the detectable  massive quantum hairs. 
   
   {\bf Acknowledgments}

Special thanks are to S. Dimopoulos,  G. Gabadadze,  N. Kaloper,  J. March-Russell  and S. Radjbar-Daemi for the valuable ongoing discussions  and suggestions. We also thank M. Bianchi  and M. Redi for comments on high spin fields. 
The work  is supported in part  by David and Lucile  Packard Foundation Fellowship for  Science and Engineering, and by NSF grant PHY-0245068. 

We also thank  Galileo Galilei Institute for Theoretical Physics  and INFN, and  Institut  des Hautes \'Etudes Scientifiques
for the hospitality 
 during the completion of this work.

\vspace{0.5cm}   


\end{document}